\documentstyle[12pt,moriondpp,epsfig]{article}

\def\mathfont#1{\ifmmode{#1}\else{$#1$}\fi}

\def\la{\mathrel{\hbox{\rlap{\hbox{\lower4pt\hbox{$\sim$}}}\hbox{$<$}}}}
\def\ga{\mathrel{\hbox{\rlap{\hbox{\lower4pt\hbox{$\sim$}}}\hbox{$>$}}}}

\def\msun{\mathfont{{\rm M}_\odot}}

\def\zform{\mathfont{z_{\rm F}}}

\begin{document}

\null

{\vskip -2.3truecm  \hskip 3.5truecm
  \vtop{\hsize=12.truecm \hyphenpenalty=5000 \noindent
     To appear in {\it Building the Galaxies: From the Primordial
     Universe to the Present} (XIXth Moriond Astrophysics Conference)
     eds. F. Hammer, T. X. Thuan, V. Cayatte, B. Guiderdoni, \& J. T. T. Van, 1999.}}

\heading{ENERGY DISTRIBUTIONS AND THE FORMATION TIMES OF SPHEROIDAL
POPULATIONS}

\author{Robert W. O'Connell} 
\affil{University of Virginia, 
       Charlottesville, USA 22903-0818 }

\begin{moriondabstract} 

I review recent progress in exploring the formation times of spheroidal
stellar populations (elliptical galaxies and large spiral bulges) using
spectrophotometric techniques.  A quickly growing body of evidence
shows that although massive spheroids can form at early times, there
are strong environmental dependencies, and major transitions in star
formation histories and even morphologies are detectable to
surprisingly small redshifts ($z \sim 0.2$).  These features are
consistent with neither the strict monolithic collapse nor hierarchical
merging scenarios.  Restframe UV observations are a promising means
of improving our understanding of spheroid evolution. 

\end{moriondabstract}

\section{Introduction}

Unlike the highly varied stellar populations of spiral disks, the
spheroidal populations found in elliptical galaxies and the bulges of
large spiral galaxies are outwardly homogeneous.  Over 80\% of 
bright nearby spheroidal populations have similar, cool
energy distributions and smooth spatial light distributions, suggesting that
they are if not entirely quiescent at least in a phase of slow
evolution.  Since spheroids are dynamically hot systems, cool gas (from
internal or external sources) cannot survive long within them.  It
will be quickly transformed either into stars or into a high
temperature atmosphere which is resistant to further star formation.
Spheroidal populations therefore reflect physical conditions in the
distant past better than any other galaxy populations, and they have
always been appealing as a means to test the early evolutionary
history of galaxies.

Enough information on spheroidal populations at high redshifts is now
becoming available to attempt to test scenarios for galaxy formation.
Two extreme pictures have been under discussion for many years.  In
{\it monolithic collapse}, spheroids represent large initial
perturbations and were the first massive stellar systems to form.  This
occurred at high redshift ($z \ga 5$-10) in roughly
synchronized, intense, short-lived bursts of star formation.  Spheroids
evolved predominantly in isolation, and their masses did not change
much after the initial collapse. A strong prediction of this picture
is that evolution of the luminosities, luminosity functions, and
spectra of spheroids should be strictly passive (i.e.\ without
significant star formation) at lower redshifts.

In {\it hierarchical merging}, spheroids are less fundamental.  They
grow through stochastic assembly from small amplitude seed systems, often
disks.  Spheroid formation continued for an extended period to
relatively low redshifts, involving strong environmental interactions
between fragments and punctuated by star formation bursts if these
contain any gas.  Masses of spheroids increased with time.  In this
picture, the approach to a state of passive evolution will depend on
the environment, happening earlier in denser regions.  Models for
regions typical of the present-day, low-density field tend to predict
that strong evolution should be detectable at relatively low redshifts
$z \sim 0.5$--1.  

In this review I discuss recent progress made in exploring the
formation times of spheroidal populations using their spectral
energy distributions.  Important questions include: what is the
earliest verifiable formation time for a spheroidal population?
what is the range of formation times? and what is a typical
star formation history?

\section{Spectrophotometric Tests of Formation Time}

A comparison of the integrated spectral energy distributions (SEDs) of
a young (30 Myr) and old (10 Gyr) population is shown in Figure 1.
The SED of the younger system rises sharply to shorter wavelengths.
It is relatively smooth, showing (at this resolution) mainly the
stronger hydrogen absorption lines and continuum discontinuities.  In
the units plotted, the older SED is flat above 4500\AA\ but exhibits a
``blue precipice'' with increasing slope at shorter wavelengths with
very strong absorption features, mostly blends of metallic lines.  The
younger SED is produced almost entirely by main sequence stars at
wavelengths below 8000 \AA.  In the older SED, red giant branch stars
dominate for wavelengths $>$ 5000 \AA.  Since the temperature and
luminosity function of the RGB evolves only slowly, so does this part
of the older SED.  However, for $\lambda < 4000\,$ \AA, over 75\% of
the light in the old SED comes from the main sequence (near 1 \msun),
which evolves faster. 

\begin{figure}

\centerline{\epsfig{file=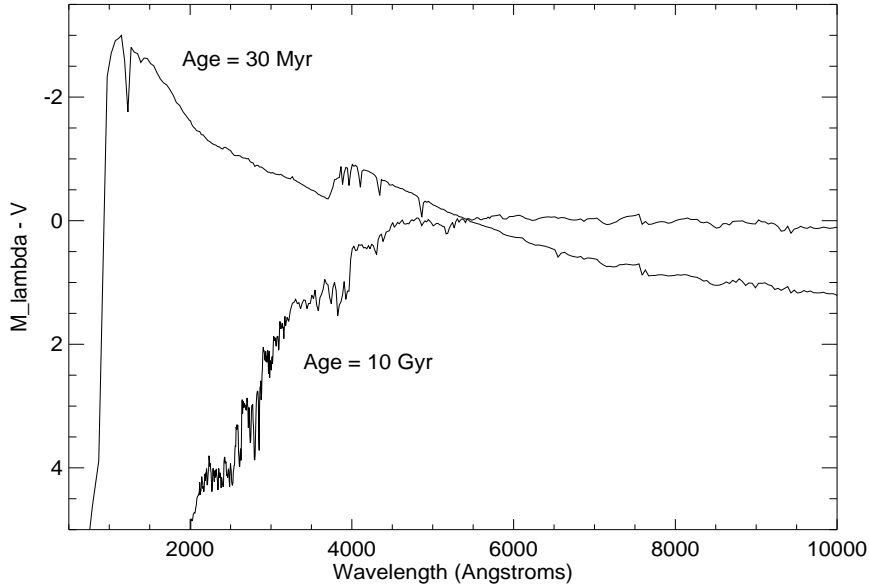,height=3.3in,width=4.8in}}

\vspace{4pt}
\caption{Comparison of energy distributions for a young and
old stellar population.   SED units are flux per unit wavelength
converted to magnitudes.  Normalized here at 5500 \AA, but the young
population is 140 times brighter per unit mass.  Taken from the
synthesis models of \cite{bc}.  See text for further details.  }
\end{figure}

The characteristic flat spectrum with a blue dropoff is a readily
identifiable  signature of any population older than
$\sim 0.5$ Gyr and has been widely used in searches for evolved
systems at high redshifts.  (Dropoffs in younger SEDs caused by the
Lyman discontinuity at 912 \AA\ or the Lyman forest below 1216 \AA\ can
look qualitatively similar in low S/N data but are usually
distinguishable by the slope longward of the cutoff.)  The pronounced
structure of the blue precipice also permits good determinations of
photometric redshifts from low resolution data.  

The spectra in Fig.\ 1 are normalized at the V-band and do not reflect
the difference in light-to-mass ratio for different ages.  Because of
the rapid decline in brightness as the main sequence burns down, the
older spectrum would be 140 times fainter than the younger one for a
given mass of stars with a normal initial mass function.  By the same
token, however, the cool SEDs of older systems permit sensitive
detection of tiny amounts of recent star formation at UV wavelengths.
Referring to Fig.~1, we see that a 30 Myr-old SED would be easily
detectable below 2000 \AA\ in the spectrum of a predominantly old
system even if it contributed only 1\% of the V-band light.  For a
normal IMF, a young component of this amplitude would contain only
0.01\% of the galaxy's mass.

The strengths and weaknesses of testing galaxy formation times using
their SEDs are illustrated in Figure 2.  This shows a hypothetical
distribution of galaxy {\it light-weighted ages} (at an arbitrary
wavelength) as a function of redshift.  Two intrinsic limitations of
SED methods are obvious:  first, SED methods cannot probe the {\it
mass}-weighted ages in which we are most interested.  Second, time
scales derived from SED analysis refer only to star formation,
not to possible dissipationless assembly of systems at later times.  

\begin{figure}

\centerline{\epsfig{file=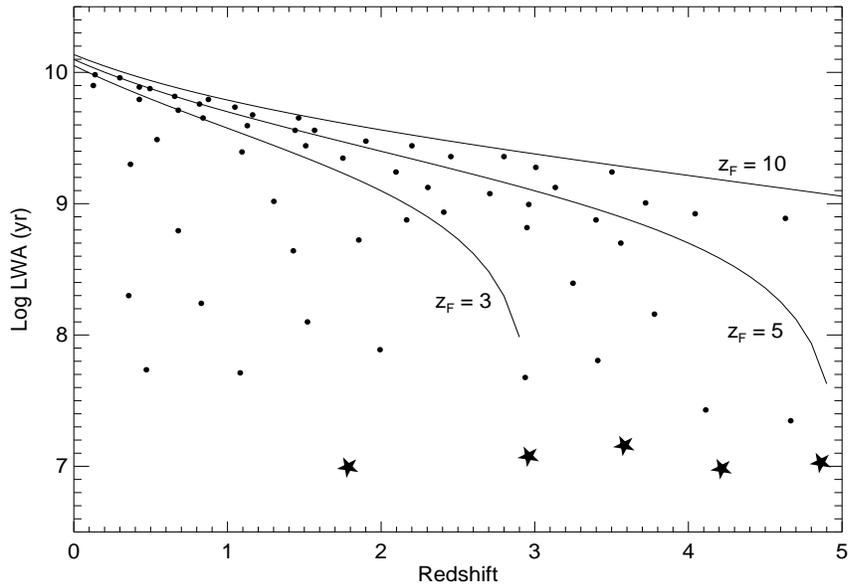,height=3.3in,width=4.8in}}

\vspace{4pt}
\caption{Schematic distribution of galaxy light-weighted ages (LWA) with
redshift in a hypothetical field sample.  Trajectories for
single-burst systems with 3 different formation redshifts ($z_{\rm
F}$) are shown.  Stars represent initial burst systems, and dots
represent evolved objects.  Most objects fall between the upper (red)
and lower (blue) envelopes because they contain multiple generations.
Ages assume H$_0$ = 65, q$_0$ = 0.}

\end{figure}

A less obvious difficulty is that younger stars dominate the light of
a galaxy if there has been significant recent star formation,
regardless of the actual age of the oldest populations.  Because of
this swamping of the older light, only poor constraints can be placed
on earlier star formation in systems containing multiple generations
of stars.  Unfortunately, this limitation implies that most of the
region inside the boundaries shown in Fig.\ 2 is not useful in
constraining the early history of galaxies.

The best tests of formation times lie at the upper and lower envelopes
in Figure 2.  First, one can search directly for starbursting
primordial systems at or near their formation redshifts ($z_{\rm F}$)
in the lower ``blue envelope.''  This is
challenging because of the (possible) large distance moduli ($\ga 50$)
involved and the shift of the stellar spectrum to the near-infrared,
where the ground-based sky is bright and one requires cooled
telescopes in space.  Heavy internal extinction by dust is also
expected to be common.  Deep IR observations of starlight or
IR/sub-mm observations of the ISM, now becoming feasible,
can identify distant starbursts, though determining the duration of
the burst or whether there were earlier bursts is difficult. 

Alternatively, one can study the least active systems near the upper ``red
envelope'' and {\it infer} \zform\ from their SEDs.  The advantages
are that this can done at lower redshifts, so much better S/N is
possible, and that infrared telescopes are not required.  Dust is also
likely to be less of a problem.  One can probe formation times by
analyzing SEDs of individual galaxies or statistical samples in
various slices through (magnitude, color, redshift) space.

The main difficulty with the age-dating of older SEDs is that they
evolve {\it slowly}.  This is because the luminosity and temperature
of main sequence stars are power laws in mass (or age) and therefore
change more slowly as age increases.  Models show that
spectrophotometric properties such as continuum colors (expressed in
magnitudes) change logarithmically:  ${\rm Color} \sim a + b \log t$,
where $t$ is the age.  (Hence, SED changes are proportional to 
the ordinate in Fig.~2.)  The coefficient $b$, which determines the
evolutionary rate, is given in the following table for several
restframe colors:  

\medskip

\begin{center}

\begin{tabular}{lc}

Color & $b$\\

\hline

B-V & 0.3 \\

U-V & 0.8 \\

U-J & 1.4 \\

2500-V & 2.0
\end{tabular}

\end{center}

\noindent $b$ increases at shorter wavelengths and as the wavelength
baseline increases.  The logarithmic color dependence means that age
resolution is given by $t/\delta t \sim b/ 2.3\, \delta{\rm C}$ and is
directly proportional to $b$ for a given observational precision.
This relation translates to stiff demands on data quality.  For
instance, the ``3-$\sigma$'' age interval derived from restframe U--V
data with an unusually good precision of $\sigma_{\rm obs} = 0.05$ mag
is $\pm 43$\%.  

There have been some attempts to estimate absolute ages from color
scatter ($\delta$C), but we see that color scatter constrains only the
fractional range in age, not age itself (unless we have independent
information on $\delta t$).  


\section{Spheroids in Rich Clusters}

Recent studies have shown that spheroids can form at $z > 1$ in
the dense environment of rich clusters of galaxies.  The red envelope
is in place in clusters at redshifts at least up to $z = 0.9$, or about
50\% of the age of the universe \cite{aecc}, \cite{esdco}, \cite{sed}.
The envelope is composed largely of E or S0 systems at HST resolution.
The slope of the color-magnitude relation (driven by the
metallicity-luminosity correlation) is preserved, and the envelope
shows impressively small scatter in the photometric-dynamical
correlations of the fundamental plane, e.g.\ \cite{vdf}, and also in
color ($\sigma [U-V]_{\rm rest} \la 0.1$).  However, mean colors are
offset to bluer values with respect to nearby clusters such as Coma,
with $\Delta (U-J)_{\rm rest}
\sim -0.5$.  The color trend is consistent with passive evolution.
The age of these least active systems at the observed epoch is $\sim
3-5$ Gyr, implying $z_{\rm F} \ga 1.5-3$ and present-day ages for
their descendents of $\ga 9$ Gyr (H$_0$ = 65).  Even though the limits
on color scatter are small, they imply a significant range in age at
the observed epoch ($z \sim 0.7$) of $\sigma(t)/t \sim 0.3$, which
means that the max/min age ratio for the most homogeneous 68\% of the
sample is a factor of 1.7.  The available data therefore permit a
formation epoch extended over several Gyrs within and among rich
clusters.  Whether this involved mergers or simple collapse cannot be
determined yet, nor can continuous production of some fraction
($\la 25$\%) of the E galaxies through interactions be excluded.  

The populations of most E galaxies in clusters therefore probably form
early and typically suffer only minor pollution from star forming
events at $z < 1$.  However, the situation is very different
regarding the spheroids of cluster S0 galaxies.  HST observations show
the {\it mixture} of morphological types appears to differ
substantially from that in local clusters even at redshifts as low as
$z \sim 0.3$.  There are many more spirals (Sa-Sdm/Irr) and many fewer
S0's (factors of 2-4) than in nearby rich clusters \cite{cbses}
\cite{docse}.  It is the higher incidence of spirals which produces
the ``Butcher-Oemler effect'' \cite{bo} on color statistics.  A
dominant evolutionary process in clusters during the past 5 Gyr has
therefore apparently been the transformation of star-forming spirals
into quiescent S0's.  Good evidence of rapid spiral/S0 evolution is
present in radial gradients of S0 colors within clusters \cite{vd98}
and in the spectra of cluster members, roughly 30\% of which have
suffered major changes in their star formation rates (bursts or
quenching) within the preceding 2 Gyr \cite{b96} \cite{cbses}.
Careful spectral surveys have shown that similar activity continues at
a low level to the present epoch in some clusters \cite{cr}.  The data
suggest that several processes, including infall, galaxy mergers,
subcluster mergers, small accretion events, and ram-pressure
stripping, are important in morphological transformations.  One
implication is that S0 spheroids in nearby clusters are considerably
younger on average than their E neighbors.  This interpretation must
be reconciled with the well-known result that nearby cluster E's and
S0's have very similar photometric properties; one possibility is that
the photometric tests made so far are insensitive to age differentials
of 30-50\% for old populations.  

\section{Spheroids in the Field}

It is somewhat misleading to contrast rich clusters with {\it the
field}, since the latter comprises a very wide range of environments,
and these can differ substantially from the local region which we
define as ``normal.'' Large fluctuations mean that pencil-beam surveys
are always statistically insufficient.  Strong selection effects
operate to bias samples as regards depth, luminosity, star formation
histories, morphology, and other properties \cite{kk} \cite{kr}.

The most conspicuous evolutionary feature in the field is
the ``blue excess'' in the faint galaxy counts,  reviewed in
\cite{ellis97} and \cite{kk}.  Although it was initially thought this
might be associated with the formation of massive spheroids at high
redshifts, instead it appears to be produced by moderate luminosity,
moderate redshift, irregular systems, which disappear at lower
redshifts.  These may evolve into spheroidal systems, but probably not
luminous ones.  

There is no doubt that old spheroidal systems exist in the field at
high redshifts.  The best example is the radio galaxy LBDS 53W091,
which appears to be a bona-fide quiescent elliptical at a redshift $z
= 1.55$.  From restframe UV spectra, Spinrad et al.\
\cite{sp97} obtained an age of $\ga 3.5$ Gyr, implying $z_{\rm F} >
3-5$.  This age is large enough that it significantly constrains
cosmology, here requiring $\Omega < 0.4$ if H$_0 > 65$.  This
demonstrates the potential power of ``red envelope'' analysis (even if
there is debate over the appropriate limiting age value
\cite{heap98}).  Other examples of $z > 1$ red spheroids have 
recently been found in radio samples \cite{pea98} and deep HST NICMOS
observations, e.g.\   \cite{benitez} and \cite{stiavelli}.

However, the global history of spheroids in the field has yet to be
established.  Considering the selection effects and variety of
environments, it is not surprising that there is considerably more
controversy here than for clusters.  Mutually contradictory claims are
the rule.  Samples of putative spheroids in the field can be selected
on the basis of (red) color or, since the 1993 repair of HST, high
resolution morphologies and light profiles.  Although morphology is
generally considered a stronger criterion, nonetheless there are good
nearby examples of systems whose apparent E/S0 morphology is a product
of low resolution \cite{crd}, and such limitations on classification
must affect the higher redshift samples.  

Several recent studies of morphologically-selected (by HST) E/S0
galaxies conclude that there is no strong evolution in the number
density of such objects to $z \sim 1$ \cite{im} \cite{menan}
\cite{schade99}.  These results are still statistically marginal, and
a much larger sample with confirmed redshifts is desirable, but they
appear to exclude mergers at $z < 1$ as the main channel for
production of field ellipticals.  

Interestingly, however, these and other studies also appear to exclude
the monolithic collapse picture for spheroid production, at least in
the strict form which requires quiescent post-burst behavior.  The
evidence is of two kinds.  First, several studies find a deficit in
color-selected samples in the number of ``ultrared'' (e.g. I--K $> 4$)
objects which would correspond to old, quiescent populations at
redshifts $z > 1$  \cite{b99} \cite{kcw} \cite{zf}.  It is not clear
whether these results are statistically inconsistent with the (as yet)
small sample of red spheroids known at higher $z$.

Second, many of the morphologically-selected E/S0's and spiral bulges
at redshifts up to $z \sim 1$ show evidence of significant star
formation in the preceding 2 Gyr \cite{ab99} \cite{franc98}
\cite{menan} \cite{schade99}.  Schade et al.\ \cite{schade99} find
that 30\% of the E's have strong [O II] emission, unlike local
samples, and Abraham et al.\ \cite{ab99} find color evidence of recent
activity in a comparable percentage of HDF spheroids.  Mananteau et
al.\ \cite{menan} find that their spheroidal sample mixes in color
with spiral and irregular galaxies and that only a small fraction has
properties consistent with single bursts and $z_{\rm F} \ga 3$.  Most
studies argue that the bluer colors are associated with late,
small-amplitude (5-25\% of the mass) bursts of star formation rather
than the decay of the initial burst.  However, the relatively brief
lifetime of color disturbances from bursts coupled with the large
fraction of objects showing them implies that, statistically, most
local field spheroids should have had significant star forming
activity in the last 3-10 Gyr.  This is consistent with evidence for a
wide range of star-forming histories among local spheroids
\cite{faber95} \cite{oc94} \cite{rbcest}. 

\section{UV Probes of Spheroidal Populations}

Most of the studies discussed here are based on observations of older
populations at wavelengths longer than 3300 \AA\ in the restframe.
There will be considerable advantages in pushing to shorter
wavelengths.  As a glance at Fig.~1 shows, most of the information in
the SED of a stellar population is to be found at rest wavelengths
below 4000 \AA.  This is true both of age indicators (for which the
$b$ parameter rapidly increases below 3500 \AA) and metal abundance
indicators (by virtue of strong absorption features of Ca, CN, NH, Mg,
and Fe and general background blanketing).  Empirical mid-UV
(2000-3200 \AA) spectra of globular clusters and spheroidal galaxies
demonstrate great sensitivity to population parameters \cite{ponder}
\cite{rd99}.  Increasingly realistic UV spectral synthesis models 
relevant for galaxy SED analysis are appearing
\cite{bc96} \cite{dor99} \cite{heap98} \cite{frv97} \cite{sp97}.  The
highest potential sensitivity is found in the ``UV upturn'' component,
present at wavelengths below 2000 \AA\ in all local spheroids observed
to date \cite{oc99}.  This is produced by small-envelope, low-mass,
extreme horizontal branch stars and their descendents.  Simple models
predict a sudden appearance of the upturn component as the
population's turnoff mass drops below a critical threshold at an age
of $\sim 4$-8 Gyr.  A recent detection of far-UV radiation in 4 E
galaxies in the cluster A370 at $z = 0.38$ has been made by Brown et
al.\ \cite{brown98}.  If this is the upturn component, it implies a
high formation redshift ($z_{\rm F} > 4$) in the context of existing
models.  However, serious uncertainties in modeling giant branch mass
loss and helium enrichment, as well as the possibility of
contamination from young populations, render any conclusion
premature.

\section{Conclusion}

The remarkable profusion of new information on distant galaxies,
especially ``red envelope'' systems, offers tantalizing if not
conclusive insights into the basic processes of galaxy formation.  The
monolithic collapse model is an easier target because of its definite
predictions.  In its extended form, in which star formation in all
galaxies begins intensely and synchronously at high redshift and then
declines in smooth exponentials depending only on galaxy type (e.g.\
\cite{tins}), it has been the traditional foundation for interpreting
galaxy populations for 40 years.  This model is almost certainly
wrong.  There is no evidence for a unique, well-defined epoch of
galaxy formation.  Instead, galaxy evolution is accelerated in denser
environments, and major transitions in star formation histories and
even morphologies are detectable to surprisingly small redshifts ($z
\sim 0.2$).  These features are consistent with the hierarchical
models.  However, there is also good evidence, both in rich clusters
and the field, that massive spheroidal systems can form at rather
early times ($z_{\rm F} \ga 3$--5), which was not anticipated in the
standard hierarchical models.  Improvements in understanding and
modeling UV spectra of old populations promise much better sensitivity
to age and abundance.


%
\vfill

\end{document}